\documentstyle[aps,prl,epsf]{revtex} \twocolumn

\begin{document}
\title{Output of a pulsed atom laser}
\author{H. Steck, M. Naraschewski, and H. Wallis}
\address{Max-Planck-Institut f\"ur Quantenoptik, 
	 Hans-Kopfermann-Stra{\ss}e 1,
	 D-85748~Garching, Germany \\
          and \\
Sektion Physik, Ludwig-Maximilians-Universit\"at M\"unchen,  
Theresienstra{\ss}e 37, 
D-80333 M\"unchen, Germany}
\date {August 8, 1997}
\maketitle

\begin{abstract}
We study the output properties of a pulsed atom laser consisting of an
interacting Bose-Einstein condensate (BEC)
in a magnetic trap and an additional rf field 
transferring atoms to an untrapped Zeeman sublevel.
For  weak output coupling  
we calculate the dynamics of the decaying condensate population, of its
chemical potential and the velocity of the output atoms analytically.
\end{abstract}


\narrowtext

The  experimental breakthrough to Bose-Einstein condensation  
with small numbers of atoms in magnetic traps
\cite{AND95} has raised much interest
in the properties of mesoscopic quantum gases.
Bose-Einstein condensates with atoms 
in a single magnetic sublevel have been studied 
experimentally and theoretically.
The recent experimental and theoretical investigations 
of interference between two independent 
Bose-Einstein condensates convincingly proved their   
macroscopic coherence \cite{KET97,ROE97}.
Moreover, the laser-like  coherence of the atoms
is preserved in the presence 
of a matter-wave splitter based on rf-transitions 
pumping the atoms into  
untrapped magnetic sublevels  \cite{MEW97}. These states are either
strong-field seeking or have no magnetic moment at all, and 
leave the trap. Alternatively, optical Raman transitions can be used for
the transfer \cite{MOY97,NAR97}. Such schemes provide 
 controllable output couplers for coherent atom lasers.

In  analogy to a laser one can distinguish between a cw laser, based on 
continuous refilling of the condensate, and
 a  pulsed atom laser where the condensate is periodically 
refilled and slowly released, similar to
\cite{MEW97}.
Whereas a continuous wave atom laser has been studied 
only theoretically \cite{NAR97,atomlaser1,atomlaser2},
current Bose-Einstein condensation experiments are 
limited to the pulsed mode of operation. 
The closest approximation of a cw atom laser by a pulsed one
can be reached in the limit of a weak coupling rf field. In
this case we are able to describe the decay of the trapped condensate
and its energy width analytically. Previous calculations addressed 
the opposite limit of strong coupling by numerical calculations 
\cite{BAL97} or neglected the important 
influence of atom-atom interactions \cite{HOP97}.

The output coupler consists of a
monochromatic resonant rf field of frequency $\omega_{\rm rf}$ 
transferring $^{\rm 23}$Na atoms in the $F=1$ hyperfine state from
the trapped $m=-1$ into the untrapped $m=0$ and the repelled
$m=1$ magnetic sublevels. 
For simplicity an isotropic harmonic trap potential
$V_{-1}({\bf r})= V_{\rm off}+ M \omega_{\rm T}^2 {\bf r}^2/2$,
$V_{+1}({\bf r})=-V_{-1}({\bf r})$
and $V_0({\bf r})\equiv0$ are assumed while
effects of gravity are neglected. 

The three  coupled coherent matter waves 
are described by a three-component Gross-Pitaevskii equation (GPE) 
with resonant excitation in rotating wave approximation
first studied for a generic two-level system in Ref.\cite{BAL97}.

In the following we adopt the 
point of view of {\em spontaneously} broken gauge symmetry for 
a Bose gas initially at zero temperature. 
The system of equations  for  the macroscopic wave funtion  $\tilde \psi_m (t) 
= e^{ -i m \omega_{\rm rf} t} \langle \hat \psi_m (t) \rangle  $ 
in rotating wave approximation  for $m,m'\in \{-1,0,+1\}$ now  reads
\begin{eqnarray} 
i\hbar\frac{\partial }{\partial t} \tilde \psi_{m}({\bf r},t)
\!\!& =&\nonumber \\
        &&\!\!\!\!\!\!\!\!\!\!\!\!\!\!
          \!\!\!\!\!\!\!\!\!\!\!\!\!\!\!\!\!\!\!\!\! 
 \left(-\frac{\hbar^2 \nabla^2}{2M}+
	V_{m}({\bf r}) +\hbar m \omega_{\rm rf} + 
         U ||\tilde \psi( {\bf r},t)||^2 \right )
        \tilde \psi_{m}({\bf r},t) \nonumber\\ 
        &&\!\!\!\!\!\!\!\!\!\!\!\!\!\!
          \!\!\!\!\!\!\!\!\!\!\!\!\!\!\!\!\!\!\!\!\!
 + \hbar\Omega\,\sum_{m'}
        (\delta_{m,m'+1}+\delta_{m,m'-1})\tilde \psi_{m'}({\bf r},t) .
\label{3GPE}
\end{eqnarray}
Here, we have replaced the atomic density by the modulus of the wave function
\begin{equation}
\langle \hat n({\bf r},t) \rangle = ||\tilde \psi( {\bf r},t)||^2 =  
\sum_m | \tilde \psi_m( {\bf r},t)|^2\,.
\end{equation}
The coupling constant
 $\hbar \Omega=g \mu_{\rm Bohr} |B| /\sqrt{2}$
refers to the Rabi frequency due to the rf field. 
At zero magnetic field 
the symmetrized s-wave 
scattering matrix elements $U_{mm'}=4 \pi\hbar^2 a_{mm'}/M$ 
for an elastic collision
of a pair of atoms in the sublevels $m,m'\in \{-1,0,+1\}$ 
 are all nearly equal to
$\overline a = 53a_0 $, ($a_0$ is the Bohr radius),    
according to preliminary 
calculations of E.~Tiesinga and P.~S.~Julienne \cite{TIE97}.
Since   the steady state 
operation depends mainly on the initial condensate mean field,
we assume in the following a diagonal scattering matrix $a_{mm'} = \delta_{mm'} 
\overline a$ for simplicity. 
Consequently, the Hartree mean field potential for 
each spin component is equal to the total atom density 
$\langle \hat n({\bf r},t) \rangle$ multiplied 
by $U=4 \pi\hbar^2 \overline a/M$.

The initial condition is chosen as the solution of the 
stationary GPE for the trapped condensate in the 
absence of the rf field, i.e $\Omega=0$ in Eq.~(\ref{3GPE}). 
In the Thomas-Fermi approximation it reads
\begin{equation}
|\tilde \psi_{-1}({\bf r},0)|^2 = \max \left [ 
\frac{\mu+V_{\rm off}-V_{-1}({\bf r})}{U}, 0 \right ].
\label{TFcond}\end{equation}

\begin{figure}
\begin{center}
\leavevmode
\epsfxsize=0.5\textwidth
   \epsffile{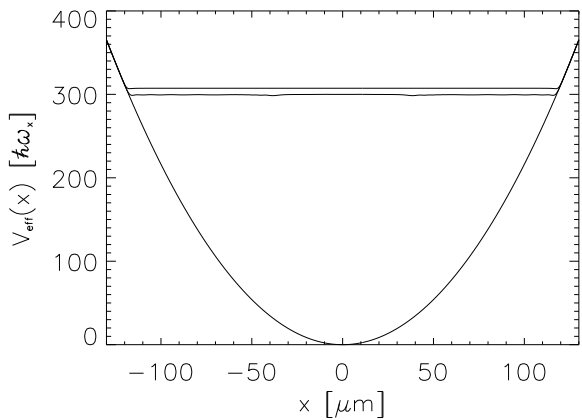}
\end{center}
\caption{\label{F_decay}
Uniform decay of the condensate density in terms of the effective potential
$V_{-1}({\bf r})-V_{\rm off}+ U|\tilde \psi_{-1}({\bf r})|^2$ for 
small field strength $\Omega$ = 12~s$^{-1}$, $N_0=5\times10^6$ and 
$\Delta(0)$ =
3100~s$^{-1}$ in a one-dimensional situation.
 The lower density is reached after 670~ms. 
  }
\end{figure}

For a small coupling strength ($\Omega\ll\omega_{\rm T}$) the
process of atoms leaking out of the resonance points
is faster than the Rabi oscillations. Therefore 
the coupling into state $m=+1$ can be neglected, since
it is proportional to $\Omega^4$. In the following
only the states $m=-1$ and $m=0$ are considered.
After switching on the coupling due to the rf field,
initial oscillations   die out quickly, because the untrapped
atoms leak out of the trap within less than one
Rabi cycle.  Other condensate atoms move into the resonance area
replacing the leaving ones.
Eventually a  quasi-stationary state is reached, i.e.  
the $m=-1$ condensate wave function decays slowly without oscillations while the 
atoms coupled out of the condensate are expelled due to the mean-field potential
 and form a steady current.
A numerical solution of the two-component GPE shows the uniform decay
of the trapped condensate (cf. Fig.~\ref{F_decay}).
The sum of external and mean field potentials 
  is in a good approximation spatially independent
inside the condensate so that a description of the decay in terms of
the chemical potential $\mu(t)$ is appropriate.
 In the following we present an analytical calculation of
the time dependent output intensity in this quasi-stationary
regime, assuming a three dimensional isotropic 
harmonic trap.

We first calculate  
the rate $\Gamma$ of transitions from the
condensate into the output. In the spirit of the 
Thomas-Fermi approximation we neglect the kinetic energy
in a two-component GPE corresponding 
to Eq. (\ref{3GPE}) and solve for the output  
density distribution
\begin{equation}
|\tilde \psi_0( r,t)|^2=\frac{4\Omega^2\sin^2[\frac{1}{2}
\sqrt{\Delta^2(r)+4\Omega^2} \, t] }{\Delta^2(r)+4\Omega^2}
\,|\tilde \psi_{-1}( r,0)|^2\,.
\end{equation}
The maximum amplitude of the Rabi oscillations is located at $r_{\rm res}$
 determined  by the resonance condition 
\begin{equation}
 \hbar \Delta(r_{\rm res}) =0
\label{reson} \end{equation}
where  $\hbar \Delta(r)  = \hbar \omega_{\rm rf} - V_{-1}(r)$.
Thus, the main contribution to the output coupling stems from
a small shell around that resonance radius. 

The time-derivative of the density gives the density transition rate.
The total transition rate 
is obtained by integrating the position dependent rate over the condensate 
volume. 
The transition rate is negligible outside a minute resonance shell 
and strongly peaked within that shell. 
Expanding the position dependent detuning 
$\Delta(r)$ to first order 
 one obtains the time dependent transition probability 
\begin{eqnarray}
\Gamma(t) &\equiv& \frac{1}{|\tilde \psi_{-1}( r_{\rm res},0)|^2}\int^{\infty}_0 
4 \pi r^2\frac{\partial}{\partial t }|\tilde \psi_0(r,t)|^2 dr \nonumber\\
&&\approx 4\pi r_{\rm res}^2
\int  _{-\infty} ^{\infty}  2 \Omega^2
\frac{
  \sin 
[( \sqrt {{\Delta'(r_{\rm res})}^2 r^2 + 4 \Omega^2  }\, t] }
{\sqrt { {\Delta'(r_{\rm res})}^2 r^2 + 4 \Omega^2 } } dr \nonumber\\
&&=  8\pi^2r_{\rm res}^2 \frac{ \Omega^2}
{\Delta'(r_{\rm res}) }   J_0 ( 2 \Omega t ).
\end{eqnarray} 
The calculation of the above rate $\Gamma(t)$ does not account for the
losses due to the leaving atoms. However, a rate equation allowing for these
losses can be derived in the limit of weak coupling by using the 
perturbational rate ($t\ll \pi/\Omega$)

\begin{equation}
\Gamma = 8\pi^2\hbar\Omega^2\frac{\sqrt{2\hbar\Delta(0)}}
{(M\omega_{\rm T}^2)^{3/2}}
\label{rate} \end{equation}
with $\Delta^\prime (r_{\rm res}) = 2 \Delta(0)/r_{\rm res}$.
 
Due to the spatial localization 
of the output coupling, the decay of the condensate population
\begin{equation}
N(t)\equiv\int d^3r  |\tilde \psi_{-1}( r,t)|^2  \,\,,\,\, N(0)=N_0
\end{equation}
depends solely on the density 
of the atoms around the resonance shell with radius $r_{\rm res}$
\begin{equation}
\frac{dN(t)}{d t} 
=- \Gamma \,|\tilde \psi_{-1}(r_{\rm res},t)|^2\,.
\label{decay} 
\end{equation}
In the quasi-stationary regime we assume
the shape of the condensate density being
 equal to the Thomas-Fermi solution of the 
stationary GPE (cf. Fig.~\ref{F_decay})  with a slowly varying 
atom number $N(t)$,
\begin{equation}
N(t) =\frac{4 \pi}{15 {U}} [2\mu(t)]^{5/2}/
(M \omega_{\rm T}^2)^{3/2}. \label{thomasfermi} 
\end{equation}
The condensate density at the resonance points is then given by
\begin{equation}
|\tilde \psi_{-1}( r_{\rm res},t)|^2=[\mu(t) -\hbar \Delta(0)]/{U}.
\end{equation}
Inserting this into the decay law (\ref{decay}) 
we obtain a nonlinear differential
equation for the decay of the chemical potential
\begin{equation}
\frac{d}{d t}\mu + \alpha 
\,\frac{\mu-\hbar\Delta(0)}{\mu^{3/2}}=0,
\label{muoft}
\end{equation} where
$ \alpha = 3\Gamma (M\omega_{\rm T}^2)^{3/2}2^{-7/2}\pi^{-1}$.
Integration yields  
\begin{eqnarray} 
&&\Bigg [
2 \hbar \Delta(0) \mu^{1/2}  + \frac{2}{3} \mu^{3/2}  \nonumber\\
&&\quad  -2[\hbar \Delta(0)] ^{3/2} {\rm artanh} 
\sqrt { \mu /\hbar \Delta(0)} \Bigg ] ^{\mu(0)}_{\mu(t)} = \alpha t\,\,.  
\end{eqnarray} 

With Eq.~(\ref{thomasfermi}) this  yields additionally the time evolution for
the number of trapped atoms $N(t)$ as well as  the flux
and the velocity of the untrapped atoms. 

During the depopulation of the condensate the chemical potential 
and the spatial extension of the condensate decrease until
the resonance points lie on the surface of the shrinked 
condensate. At this point the flux out of 
the condensate vanishes, because $|\tilde \psi _{-1}(r_{\rm res},t)|^2=0$,
and the chemical potential and the 
number of atoms remaining in the trap become constant in time 
\begin{eqnarray}
\mu(\infty) &=& \hbar \Delta(0).
\end{eqnarray}

\begin{figure}
\begin{center}
\leavevmode
\epsfxsize=0.4\textwidth
   \epsffile{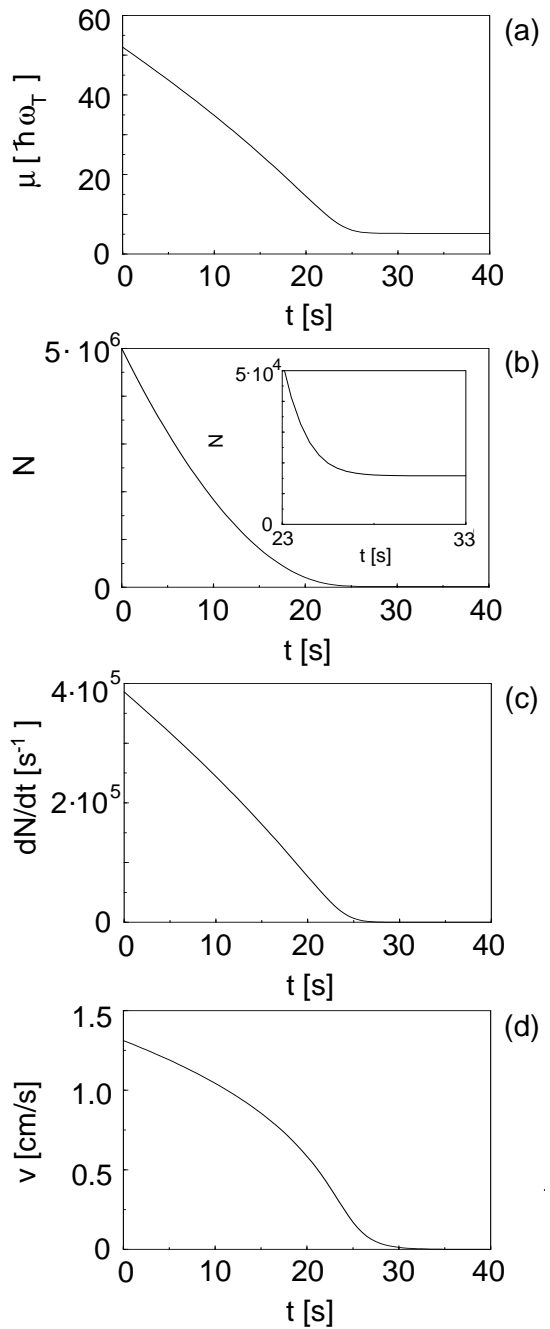}
\end{center}
\caption{\label{F_rates}
Time evolution in the Thomas-Fermi approximation:
(a) chemical potential, (b) number of particles in the 
trapped $m=-1$ state, (c) particle flux in the untrapped 
$m=0$ state, 
(d) velocity of atoms leaving the condensate.
The parameters are  $N_0=5 \times 10^6$,
$ \Delta(0)=3500$\,s$^{-1}$, $\Omega=20$\,s$^{-1}$.
  }
\end{figure}

As an example we chose a small value of $\Delta(0)$ and  
a trap frequency $\omega_{\rm T}=2 \pi \times 106$~Hz, 
the geometric mean of the values given in Ref.\cite{MEW97}. 
The resulting time evolution of the system variables according to 
the nonlinear differential equation is shown
in Figs.~\ref{F_rates}(a) and (b). The 
chemical potential $\mu(t)$ and the number of trapped atoms 
reach their steady state after roughly 25 seconds. Correspondingly
the flux and the velocity of the untrapped  atoms decrease
 to zero (cf. Figs.~\ref{F_rates}(c) and (d)). 
The corresponding calculations were also carried out in one dimension 
in order to compare the  analytical expressions 
with numerical simulations of the full coupled GPE's.
The results showed excellent agreement
 (less than 5 \% deviation).

The finite duration of the atom pulse 
ejected from the trap leads to a finite energy
width of both the condensate and the output beam that 
is called here 
 the {\em natural energy width} $\delta E$ of a pulsed atom laser, 
in analogy to the natural linewidth of spontaneously emitted photons. 
It can be calculated approximately by describing the initial stage of
the output coupling process by an exponential decay of the condensate
population
\begin{equation}
\frac{dN}{dt}(t) = -\Gamma_{\rm pop}\, [N(t)-N(\infty)].
\end{equation}
The rate $\Gamma_{\rm pop}$ results in an energy uncertainty
\begin{equation}
\delta E = \hbar\Gamma_{\rm pop} \approx 
\frac{15}{2}\pi\hbar^{5/2} \frac{\Omega^2\sqrt{\Delta(0)}}{\mu(0)^{3/2}}
\end{equation}
Whereas the transition rate $\Gamma$ does not depend 
on  $\mu$ and $N_0$, the population decay rate $\Gamma_{\rm pop}$ 
does so due to its dependence on $\tilde \psi_{-1}(r,0)$.
We thus find a dependence of the energy width on the condensate 
number 
\begin{equation}
\delta E \propto \frac{\Omega^2 \sqrt{\Delta(0)}}{N_0^{3/5}}.
\end{equation}
The natural energy width becomes narrower for weaker coupling strength $\Omega$,
i.e. a slower output coupling process.
 The same effect can be achieved by choosing
a smaller detuning $\Delta(0)$  which causes 
the sphere of the resonance points to shrink 
towards the center of the condensate. The energy width can  further be reduced 
by starting out from a condensate with a large population $N_0$.

The energy width $\delta E$ can also be understood as a velocity width
$\delta v= \delta E/(Mv)$ of the untrapped atoms leaving the condensate.
The parameters given in Fig.~\ref{F_rates} result in
$\langle v \rangle =1.31$cm/s for the velocity outside the condensate and
 $\delta v/\langle v \rangle \approx  10^{-6}$ for the relative
velocity width.

An additional energy width is imposed by the temporal decay
of the chemical potential $\mu(t)$. It leads to a decrease of the
output velocity according to $Mv(t)^2/2= \mu(t)-\hbar\Delta(0)$ thereby
implying a frequency chirp of the output beam.
 This frequency chirp can be
compensated, however, by imposing a  chirp on the 
frequency  $\omega_{\rm rf}$ of the 
rf field such that the output velocity rather than the detuning $\Delta(0)$
 becomes
 constant.

The output pulse can last for up to the order of 100 seconds
such that  causes of phase fluctuation which would also 
influence a cw-atom laser have to be considered as well.
Additional broadening of the energy width of the output beam is caused by
thermal excitations
of the condensate wavefunction and technical noise of the output coupling
mechanism such as fluctuations of the confining magnetic field.
Thus, the {\em natural energy width} calculated above has to be understood
as a lower limit. A comprehensive theoretical description of the fluctuations
of the condensate wavefunction, particularily of its phase, is beyond the
scope of this Letter.

It has been shown in \cite{NAR97} 
that the temporal decay of phase correlations 
transforms
into a decay of the spatial coherence of the output beam along the mean
classical trajectory that is being passed during the according correlation
time of the condensate phase.

At present, random variations of the
bias magnetic field at the mGauss level are
among the most important experimental limitations to the
coherence properties of the output, prior to fluctuations of the condensate
wavefunction. They might contribute to a phase diffusion of the output beam
in the 100~Hz range \cite{ESS97}.

 Pulsed atom lasers will play an important role 
in creating coherent matter waves as the 
size of the condensate may be largely increased 
in future experiments. The results of this Letter 
are based on the solutions of the coupled Gross-Pitaevskii equations 
and allow us to extract the relevant properties of the output
analytically.
The velocity of the untrapped atoms depends on the 
slowly decaying population of the trapped condensate fraction, 
leading to a slow chirp of the output frequency. 
While this frequency chirp might be compensated by a variation
of the rf-field frequency, other sources of spectral width like 
fluctuations of the confining magnetic field are surely more 
difficult to overcome.

We are grateful to P.~S.~Julienne and E.~Tiesinga for communicating
their unpublished data about the  scattering lengths 
for $F=1$ sodium atoms, and to A.~Schenzle for valuable 
discussions. Financial support by the Deutsche 
Forschungsgemeinschaft (grants Wa 727/4-1 and Wa 727/5-1) 
is acknowledged.

\begin{thebibliography}{10}

\bibitem{AND95}          % [1]
 M.\,H.\,Anderson, J.\,R.\,Ensher, M.\,R.\,Matthews, C.\,E.\,Wieman, and
E.\,A.\,Cornell, Science {\bf 269}, 198 (1995);
 K.\,B. Davis, M.-O. Mewes, M.\,R. Andrews, N.\,J.\,van Druten, D.\,S.
Durfee, D.\,M. Kurn, and W.\,Ketterle, Phys.\,Rev.\,Lett. {\bf 75}, 3969
(1995).


\bibitem{KET97} M.\,R.\,Andrews, C.\,G.\,Townsend, 
H.-J.\,Miesner, D.\,S.\,Durfee, D.\,M.\,Kurn
and W.\,Ketterle, Science {\bf 275}, 637 (1997).

\bibitem{ROE97} A.\,R\"ohrl, M.\,Naraschewski,  A.\,Schenzle, and H.\,Wallis,
 Phys.\,Rev.\,Lett. {\bf 78}, 4143 (1997).

\bibitem{MEW97} M.-O.\,Mewes, M.\,R.\,Andrews, D.\,M.\,Kurn, 
D.\,S.\,Durfee, C.\,G.\,Townsend, and
W.\,Ketterle,  Phys.\,Rev.\,Lett. 
{\bf 78}, 582 (1997).


\bibitem{MOY97} G.\,M.\,Moy, J.\,J.\,Hope, and C.\,M.\,Savage,
 Phys.\,Rev.\,A {\bf 55}, 3631 (1997) 

\bibitem{NAR97}      
 M.\,Naraschewski,  A.\,Schenzle, H.\,Wallis, Phys.Rev.A {\bf 56}, 603 
  (1997). Note that the definition of the detuning 
$\hbar \Delta = \mu + V_{\rm off} - \hbar \omega_{rf}$ in that paper
is equal to $\mu -\hbar \Delta$ here.

\bibitem{atomlaser1}
R.\,J.\,C. Spreeuw, T.\,Pfau, U.\,Janicke, and M.\,Wilkens,
Europhys.\,Lett. {\bf 32}, 469 (1996);
M.\,Olshanii, Y.\,Castin, and J.\,Dalibard, in {\em Laser Spectroscopy
XII}, edited by M.\,Inguscio, M.\,Allegrini, and A.\,Sasso (World
Scientific, Singapore, 1996), p.7;

\bibitem{atomlaser2}
H.\,M.\,Wiseman, A.\,Martins, and D.\,F.\,Walls, Quantum and Semiclassical
Optics {\bf 8}, 737 (1996);
M.\,Holland, K.\,Burnett, C.\,W.\,Gardiner, J.\,I.\,Cirac, and P.\,Zoller,
Phys.\,Rev.\,A {\bf 54}, R1757 (1996).


\bibitem{BAL97}
R.\,J.\,Ballagh, K.\,Burnett, and T.\,F.\,Scott,  Phys.\,Rev.\,Lett. 
{\bf 78}, 1607 (1997).

\bibitem{HOP97} J.\,J.\,Hope, Phys.\,Rev.\,A {\bf 55}, R2531 (1997).

\bibitem{TIE97} E. Tiesinga and P.\,S.\,Julienne, unpublished (1997).
The values are $a_{00}=51 a_0$, $a_{1,-1}=49 a_0$ and
$a_{m,m'}=53 a_0$ for the other $m,m'\in \{-1,0,+1\}$.


\bibitem{ESS97} T. Esslinger, private communication.

\end{thebibliography}
\end{document}